\documentclass[aps,pre,reprint,amsmath,amssymb,longbibliography,lengthcheck,showpacs,superscriptaddress,floatfix]{revtex4-2}
\usepackage{enumerate}
\usepackage[dvipdfmx]{graphicx}
\usepackage{color}
\usepackage[maxfloats=256]{morefloats}
\def\bra#1{\langle #1|}
\def\ket#1{|#1\rangle}

\begin{document}
%
\title{Effective medium theory for viscoelasticity of soft jammed solids}
%
\author{Hideyuki Mizuno}
\email{hideyuki.mizuno@phys.c.u-tokyo.ac.jp}
\affiliation{Graduate School of Arts and Sciences, The University of Tokyo, Tokyo 153-8902, Japan}
\author{Atsushi Ikeda}
\affiliation{Graduate School of Arts and Sciences, The University of Tokyo, Tokyo 153-8902, Japan}
\affiliation{Research Center for Complex Systems Biology, Universal Biology Institute, The University of Tokyo, Tokyo 153-8902, Japan}
%
\date{\today}
%
\begin{abstract}
The viscoelastic properties of soft jammed solids, such as foams, emulsions, and soft colloids, have been the subject of experiments, with particular interest in the anomalous viscous loss.
However, a microscopic theory to explain these experimental results is still lacking.
Here, we develop an effective medium theory that incorporates the effects of contact damping.
The theory explains experimentally observed viscoelastic properties, particularly attributing the anomalous viscous loss to marginal stability in amorphous systems.
This work establishes a microscopic theory for describing the impact of damping on soft jammed solids and their viscoelastic behaviors.
\end{abstract}
%
\maketitle
%
\section{Introduction}
Soft jammed solids are a type of amorphous material consisting of densely packed mesoscopic or macroscopic particles~\cite{Rheology,Mezzenga_2005,Bonn_2017,Nicolas_2018}.
They are commonly found in our daily lives, present in foods, pastes, cosmetics, soils, and more.
Understanding the elastic and viscous properties of these materials is important in condensed matter physics and material science, both from a fundamental and practical standpoint.
Previous experiments have extensively investigated foams, emulsions, and soft colloids using macrorheology and microrheology techniques~\cite{Mason_1995,Cohen-Addad_1998,Hebraud_2000,Gopal_2003,Besson_2008,Kropka_2010,Krishan_2010,Gupta_2012,Basu_2014,Nishizawa_2017,Conley_2019}.
These experiments measure the complex modulus $G^\ast(\omega) = G'(\omega) + i G''(\omega)$, where $i$ represents the imaginary unit and $\omega$ is the frequency of applied strain.
It has been reported that soft jammed solids exhibit characteristic frequency dependences in $G^\ast(\omega)$, which are distinct from standard elastic and fluid media like Hookean elastic bodies and Newtonian fluids.
In particular, anomalous viscous loss~\cite{Liu_1996}, characterized by the storage modulus $G' \propto \omega^0$ and the loss modulus $G'' \propto \sqrt{\omega}$, has garnered attention.

A characteristic of soft jammed solids is that constituent particles are strongly damped due to viscous forces~\cite{Durian_1995,Durian_1997}.
Recent studies have established that strong damping is integrated with vibrational properties of amorphous systems, giving rise to viscoelastic properties characteristic of soft jammed solids~\cite{Tighe_2011,Baumgarten_2017,Hara_2023,Hara2_2023,Mizuno_2024}.
Refs.~\cite{Tighe_2011,Baumgarten_2017} studied the macrorheology of an overdamped amorphous system and formulated the complex modulus based on vibrational eigenmodes and its vibrational density of states~(vDOS).
This formulation was extended for an underdamped system in Ref.~\cite{Mizuno_2024}.
Ref.~\cite{Hara_2023} delved into the microrheology experimental setup and formulated the corresponding complex modulus.
Additionally, the most recent work~\cite{Hara2_2023} revealed a direct link between anomalous viscous loss and excess vibrational states, known as boson peak, using a combination of theoretical and experimental approaches.

Building upon the progress made in understanding soft jammed solids, the present work aims to develop a microscopic theory for them.
Our approach is based on effective medium theory~(EMT)~\cite{EMT}, also known as coherent potential approximation theory.
The EMT has been used to address various problems, such as electronic energy levels in disordered metallic alloys~\cite{Yonezawa_1973}, conductance in electrical resistor networks~\cite{Kirkpatrick_1973}, and elastic constants in spring networks~\cite{Feng_1985,He_1985}.
It has also evolved into heterogeneous elasticity theory~(HET) to explain the elastic and vibrational properties of glasses~\cite{schirmacher_2006,Schirmacher_2007,Marruzzo_2013,Kohler_2013}, where the theory analyzes glasses as heterogeneous elastic media with fluctuating local elastic constants~\cite{Wagner_2011,Mizuno_2013}.
The HET accounts for several characteristic properties of glasses, including non-affine elastic deformation, boson peak, strong scattering of sound waves, and low thermal conductivity.
More recently, the HET has been used to interpret experimental data on several glasses~\cite{Pan_2021,Martin_2023}.

In addition, the EMT has been utilized in the study of jammed amorphous solids.
A simple model of the jammed solids is an assembly of frictionless spherical particles interacting through a harmonic potential~\cite{Hern_2003,Hecke_2010}:
\begin{equation}~\label{pot-simple}
\phi(r) = \frac{k}{2} \left( \sigma - r \right)^2 H(\sigma -r),
\end{equation}
where $\sigma$ represents the diameter of the particles, $k$ denotes the stiffness, and $H(x)$ is the Heaviside step function.
The static packings undergo the jamming transition at the density~$\phi_J$, becoming isostatic with the contact number per particle $z$ being equal to $z_c = 2d$.
Close to the transition, physical quantities, such as the static elastic moduli~\cite{Hern_2003,Hecke_2010,Mizuno_2016} and the characteristic frequencies in the vDOS~\cite{Silbert_2005,Wyart_2005,Wyart2_2005,Charbonneau2016,Mizuno_2017,Shimada_2020}, follow power-law scalings with the excess contact number $\delta z = z-z_c$.
To explain these jamming scaling laws, the EMT was applied to regard the jamming systems as random spring networks and solve the rigidity percolation problem in the mean-field approximation limit~\cite{Mao_2010,Wyart_2010,Degiuli_2014,DeGiuli2_2014}.

Soft jammed solids like foams, emulsions, and soft colloids can be well modeled by the harmonic potential in Eq.~(\ref{pot-simple})~\cite{Durian_1995,Durian_1997}.
This allows the application of the EMT based on random spring network models to these materials, as conducted in Refs.~\cite{Mao_2010,Wyart_2010,Degiuli_2014,DeGiuli2_2014}.
However, the theory has never been used to explain their viscoelastic properties, and a microscopic theory for them is still lacking.
Here we incorporate strong damping due to viscous forces into the EMT.
The theory explains experimentally observed viscoelastic properties, including the anomalous viscous loss.
The present work establishes a microscopic theory for describing soft jammed solids and the impact of damping on amorphous materials.

\section{Model and method}
\subsection{Random spring network model}
We examine a spring network model in $d$-dimensional space~\cite{Wyart_2010,Degiuli_2014}.
There are $N$ point particles placed at lattice sites, each with a contact number per site $z_0$.
Every pair of nearest neighbor particles is connected by a compressed spring.
The pair of particles also experiences a viscous force that opposes their relative velocity, which is known as contact damping.
We define the displacement vector of particle $i$ as $\vec{u}_i$~($i=1,\cdots,N$) and introduce a $dN$-dimensional vector $\ket{{u}} = \left[ \vec{u}_1,\cdots,\vec{u}_N\right]$.
For convenience, $\ket{i}$~($dN \times d$ matrix) is introduced to extract $\vec{u}_i$ from $\ket{u}$ via the inner product $\vec{u}_i = \bra{i} u\rangle$.
We consider the overdamped equation of motion for particles:
\begin{equation}~\label{eqm}
C \frac{d}{dt}\ket{u} = -M \ket{u} + \ket{F},
\end{equation}
where $t$ represents time, $M$ is the Hessian matrix, $C$ is the damping matrix, and $\ket{F} = \left[ \vec{F}_1,\cdots,\vec{F}_N\right]$ where $\vec{F}_i$ denotes the external force acting on particle $i$.

The Hessian matrix $M$ is~\cite{Wyart_2010,Degiuli_2014}
\begin{equation}
M = \sum_{\left<ij \right>} \ket{ij} \left[ k_{ij} \vec{n}_{ij} \otimes \vec{n}_{ij} - \frac{f_{ij}}{\sigma_0} \left( {I}_d -  \vec{n}_{ij} \otimes \vec{n}_{ij} \right) \right] \bra{ij},
\end{equation}
where $\ket{ij} = \ket{i} - \ket{j}$ is for a pair of nearest neighbor particles $ij$, $\vec{n}_{ij}$ and $\sigma_0$ are respectively a unit vector along the spring and the distance between these particles, ${I}_d$ is $d\times d$ unit matrix, and the summation $\sum_{\left<ij \right>}$ is taken over all the $N z_0/2$ pairs of particles $ij$.
$k_{ij}$ represents the stiffness of the spring connecting particles $ij$, and $f_{ij}~(>0)$ denotes the repulsive force exerted by the spring.
We define a dimensionless parameter, the prestress $e_{ij}$, as
\begin{equation}
e_{ij} = \frac{f_{ij}}{k_{ij} \sigma_{0}},
\end{equation}
and assume that spatial fluctuations of $e_{ij}$ are weak, thus $e_{ij}$ takes a typical value $e$, \textit{i.e.,} $e_{ij} \equiv e~(>0)$~\cite{Degiuli_2014}.

To introduce the connectivity~\cite{Wyart_2010,Degiuli_2014}, we assume that $k_{ij}$ follows the probability distribution given by
\begin{equation}
P(k_{ij}) = \frac{z}{z_0} \delta(k_{ij}-1) + \left(1-\frac{z}{z_0}\right) \delta(k_{ij}),
\end{equation}
where $\delta(x)$ is Dirac delta function and $z < z_0$.
Out of $Nz_0/2$ pairs of particles, $Nz/2$ pairs are connected by springs with a unit stiffness, while the rest are connected by springs with zero stiffness, meaning they are disconnected.
The minimum contact number required for mechanical stability is $z_c=2d$~\cite{Feng_1985,Mao_2010,Wyart_2010,Degiuli_2014}.
Therefore, the excess contact number compared to $z_c$, $\delta z = z-z_c~(>0)$, and the prestress $e~(>0)$ determine the stability of the system.
As explained in Ref.~\cite{Degiuli_2014} and as we will mention below, there is also a critical value $e_c$ for $e$, and the system remains stable only when $e \le e_c$.

The damping matrix $C$ for contact damping is~\cite{Tighe_2011,Baumgarten_2017,Hara_2023,Hara2_2023,Mizuno_2024}
\begin{equation} \label{ccontact}
C = \sum_{\left<ij \right>} \ket{ij} \mu k_{ij} \left[  \vec{n}_{ij} \otimes \vec{n}_{ij} + \left( {I}_d -  \vec{n}_{ij} \otimes \vec{n}_{ij} \right) \right] \bra{ij},
\end{equation}
where $\mu$ is the viscosity of viscous forces.
Note that contact damping only occurs between $Nz/2$ connected pairs of particles.

\subsection{Green function}
We perform the Fourier transform on Eq.~(\ref{eqm}) and obtain the Green function as
\begin{equation}
{G}({\omega \mu}) = \left(M - i\omega C \right)^{-1} = \widetilde{M}^{-1}. \label{eqgreen}
\end{equation}
Here, the matrix $\widetilde{M}$ is formulated as 
\begin{equation} \label{hesscon}
\widetilde{M} = \sum_{\left<ij \right>} \ket{ij} \left[ \widetilde{k}_{ij} \vec{n}_{ij} \otimes \vec{n}_{ij}  - \widetilde{e} \widetilde{k}_{ij} \left( {I}_d -  \vec{n}_{ij} \otimes \vec{n}_{ij} \right) \right] \bra{ij},
\end{equation}
where $\widetilde{k}_{ij}$ follows the probability distribution of
\begin{equation} \label{eqdis}
P\left( \widetilde{k}_{ij} \right) = \frac{z}{z_0} \delta\left( \widetilde{k}_{ij}-\widetilde{k} \right) + \left(1-\frac{z}{z_0}\right) \delta \left( \widetilde{k}_{ij} \right),
\end{equation}
and
\begin{equation} \label{eqtildeke}
\widetilde{k} = 1 - i \omega \mu, \qquad \widetilde{e} = \frac{e + i \omega \mu }{ 1 - i \omega \mu}.
\end{equation}
Since $\widetilde{M}$ depends on $\omega \mu$, ${G}(\omega \mu)$ is also a function of $\omega \mu$.

\subsection{Effective medium theory}
To analyze ${G}(\omega \mu)$ using the EMT~\cite{Wyart_2010,Degiuli_2014}, we introduce the effective Hessian matrix as
\begin{equation} \label{efhessiancon}
{M}_\text{eff} = \sum_{\left<ij \right>} \ket{ij} \left[ {k}^{\parallel} \vec{n}_{ij} \otimes \vec{n}_{ij} - \widetilde{e} {k}^{\perp} \left( {I}_d -  \vec{n}_{ij} \otimes \vec{n}_{ij} \right) \right] \bra{ij},
\end{equation}
and the corresponding effective Green function as
\begin{equation} \label{efhessiancon2}
{G}_\text{eff}(\omega \mu) = {M}_\text{eff}^{-1},
\end{equation}
where ${k}^{\parallel}={k}^{\parallel}({\omega \mu})$ and ${k}^{\perp}={k}^{\perp}({\omega \mu})$ are respectively the effective stiffnesses along directions parallel and perpendicular to the spring, both of which are functions of $\omega \mu$.
We also define the longitudinal ${G}^{\parallel}$ and transverse ${G}^{\perp}$ components of ${{G}}_\text{eff}$ as
\begin{equation} \label{efhessiancon3}
\begin{aligned}
{G}^{\parallel} &= \vec{n}_{ij} \bra{ij} {{G}}_\text{eff} \ket{ij} \vec{n}_{ij}, \\
{G}^{\perp} &= \frac{1}{d-1} \left[ \text{Tr} \bra{ij} {{G}}_\text{eff} \ket{ij} - {G}^\parallel \right].
\end{aligned}
\end{equation}

Next, we express $G({\omega \mu})$ as
\begin{equation}
{G} = {G}_\text{eff} + {G}_\text{eff} {T} {G}_\text{eff}, \label{eqgreenst2}
\end{equation}
where $T$ is the transfer matrix, formulated as
\begin{equation}
{T} = \sum_{\left<ij \right>} {T}_{\left<ij \right>} + \sum_{\left<ij \right>} \sum_{\left<kl \right> \neq \left<ij \right>} {T}_{\left<ij \right>} {G}_\text{eff} {T}_{\left<kl \right>} + \cdots, \label{eqgreenst3}
\end{equation}
with
\begin{equation}
\begin{aligned}
{T}_{\left<ij \right>} & = \ket{ij} \Bigg[ \frac{k^{\parallel}-\widetilde{k}_{ij} }{1- (k^{\parallel}-\widetilde{k}_{ij}) G^{\parallel} } \vec{n}_{ij} \otimes \vec{n}_{ij} \\
& - \frac{ \widetilde{e}k^{\perp}-\widetilde{e}\widetilde{k}_{ij} }{1 + (\widetilde{e} k^{\perp}- \widetilde{e} \widetilde{k}_{ij}) G^{\perp} } \left( I_d - \vec{n}_{ij} \otimes \vec{n}_{ij} \right) \Bigg] \bra{ij}.
\end{aligned}
\end{equation}
By averaging both sides of Eq.~(\ref{eqgreenst2}) over the probability distribution in Eq.~(\ref{eqdis}), denoted by $\left< \right> = \int d{\widetilde{k}_{ij}} P(\widetilde{k}_{ij})$, and implementing $\left< G  \right> = {G}_\text{eff}$, we get $\left< T \right> = 0$.
We then consider a single-site scattering process and obtain
\begin{equation} \label{emteqcon1over}
{G}^\parallel = \frac{ {k}^\parallel - \widetilde{k}({z}/{z_0})}{ {k}^\parallel \left( {k}^\parallel - \widetilde{k} \right)}, \quad
{G}^\perp = -\left[ \frac{ {k}^\perp - \widetilde{k}({z}/{z_0}) }{ \widetilde{e}{k}^\perp \left( {k}^\perp - \widetilde{k} \right)}\right].
\end{equation}
In addition, assuming the isotropy of $\bra{ij} {G}_\text{eff} \ket{ij} \propto {I}_d$~\cite{Degiuli_2014}, we obtain from Eqs.~(\ref{efhessiancon2}) and~(\ref{efhessiancon3})
\begin{equation} \label{emteqcon2over}
{G}^\parallel = {G}^\perp = \frac{2d}{z_0}\left[ \frac{1}{{k}^\parallel - (d-1) \widetilde{e} {k}^\perp } \right].
\end{equation}
Eqs.~(\ref{emteqcon1over}) and~(\ref{emteqcon2over}) provide closed equations for ${k}^{\parallel}$, ${k}^{\perp}$, ${G}^{\parallel}$, and ${G}^{\perp}$, all of which are functions of ${\omega \mu}$.
In the following, we consider an FCC lattice in three-dimensional space and set $d=3$ and $z_0=12$.

\section{Results}
\subsection{Complex modulus}
The macrorheology experiment measures the global modulus, which corresponds to the effective stiffness $k_\text{eff}$ in the EMT:
\begin{equation} \label{eqkeff}
k_\text{eff} = {k}^\parallel - (d-1) \widetilde{e} {k}^\perp.
\end{equation}
The complex modulus $G_{M}^{\ast} = G_M' + i G_M''$ is thus given by
\begin{equation} \label{eqgmast}
G_M' - i G_M'' = k_\text{eff}.
\end{equation}
On the other hand, the microrheology experiment measures the microscopic displacement of the probe particle when applying an oscillatory external force to it~\cite{Levine_2000,Atakhorrami_2006,Mizuno_2008,Wilson_2009}.
The response function of the probe particle in the EMT is
\begin{equation}
g = \frac{1}{d}\text{Tr} \bra{i} G_\text{eff} \ket{i} = \frac{2d}{z_0} \frac{1}{ k_\text{eff}}.
\end{equation}
Using the generalized Stokes relation~\cite{Squires_2010,Microrheology}, the complex modulus $G_{m}^{\ast} = G_m' + i G_m''$ is given by
\begin{equation} \label{eqgmist}
G_m' - i G_m'' = \frac{1}{3\pi \sigma_{pr} g} = \frac{1}{3\pi \sigma_{pr}}\frac{z_0}{2d} k_\text{eff},
\end{equation}
where $\sigma_{pr}$ is the diameter of the probe particle.
Thus, the macrorheology and microrheology measure similar complex moduli, both of which are characterized by $k_\text{eff}$.
We consider $G_{M}^{\ast}$ below.

We analytically solve Eqs.~(\ref{emteqcon1over}) and~(\ref{emteqcon2over}) by means of the asymptotic analysis~\cite{Degiuli_2014}, and obtain $G_{M}^{\ast}$ in Eq.~(\ref{eqgmast}) as
\begin{equation} \label{mainemaccon}
G_M' - i G_M'' = \left( 1 + \sqrt{ \frac{ e_c - e - i \omega \mu}{e_c} } \right)\frac{\delta z}{4da} + {o}\left(\delta z \right),
\end{equation}
where $a = (z_0-2d)/2d = 1$ and
\begin{equation} \label{eqec}
e_c = \left( \frac{1}{32d^2a}\frac{z_0}{2d} \right) \delta z^2 \propto \delta z^2.
\end{equation}
We provide details of the asymptotic analysis in Appendix~\ref{sec:asympto}.
By setting $\omega \mu = 0$, we obtain the static modulus as
\begin{equation} \label{mainstatic}
G_{M0}' =  \left( 1 + \sqrt{ \frac{ e_c-e }{e_c} } \right)\frac{\delta z}{4da} \propto \delta z.
\end{equation}
The $e_c$ is a critical value of the prestress $e$.
When $e$ is below $e_c$, $G_{M0}'$ can take a real number solution; however, there is no real number solution when $e$ exceeds $e_c$.
Thus, the system is in stable states at $e < e_c$, and at $e=e_c$, it gets in a marginally stable state, which is the boundary between stable and unstable states.

\subsection{Scaling laws in complex modulus}
Eq.~(\ref{mainemaccon}) provides the scaling laws in $G^\ast_M$ as
\begin{equation} \label{eqconscale}
\frac{G^\ast_M}{\delta z}
\propto
\left\{ 
\begin{aligned}
& 1 + \sqrt{\frac{\delta e}{e_c}} + i \frac{\omega \mu}{e_c}\sqrt{\frac{e_c}{\delta e}} & \hspace*{-7mm} \left( \frac{\omega \mu}{e_c} \ll \frac{\delta e}{e_c} \right), \\
& 1 + i \sqrt{\frac{\omega \mu}{e_c}} & \hspace*{-7mm} \left( \frac{\delta e}{e_c} \ll \frac{\omega \mu}{e_c} \ll 1 \right), \\
& \sqrt{\frac{\omega \mu}{e_c}} + i \sqrt{\frac{\omega \mu}{e_c}} & \hspace*{-7mm} \left( 1 \ll \frac{\omega \mu}{e_c} \ll \frac{1}{e_c} \right),
\end{aligned} 
\right.
\end{equation}
where $\delta e = e_c -e~(\ge 0)$.
At zero prestress $e=0$,
\begin{equation} \label{eqconscale2}
\frac{G^\ast_M}{\delta z}
\propto
\left\{ 
\begin{aligned}
& 1 + i \frac{\omega \mu}{e_c} & \left( \frac{\omega \mu}{e_c} \ll 1 \right), \\
& \sqrt{\frac{\omega \mu}{e_c}} + i \sqrt{\frac{\omega \mu}{e_c}} & \left( 1 \ll \frac{\omega \mu}{e_c} \ll \frac{1}{e_c} \right).
\end{aligned} 
\right.
\end{equation}
These scaling laws are consistent with those obtained by the previous works~\cite{Tighe_2011,Hara_2023,Mizuno_2024}.

\begin{figure}[t]
\centering
\includegraphics[width=0.475\textwidth]{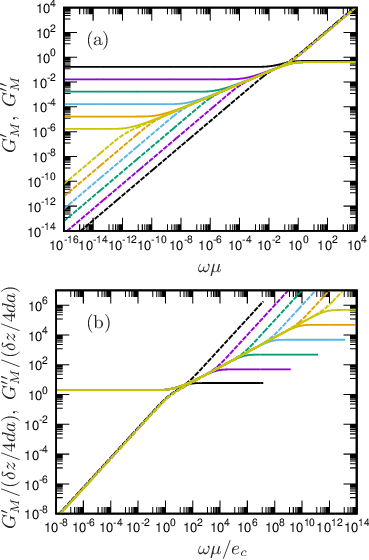}
\caption{\label{figref2con}
{Complex modulus at zero prestress.}
(a) $G_M^\ast$ is plotted as a function of $\omega \mu$, and (b) $G_M^\ast/(\delta z/4da)$ is plotted as a function of $\omega \mu/e_c$.
Solid and dashed lines represent the storage modulus $G_M'$ and the loss modulus $G_M''$, respectively.
The prestress is $e=0$.
The excess contact number is $\delta z= 10^{-5}$~(yellow), $10^{-4}$~(orange), $10^{-3}$~(cyan), $10^{-2}$~(green), $10^{-1}$~(purple), and $10^{0}$~(black).
Note that $e_c$ depends on $\delta z$ as $e_c \propto \delta z^2$.
}
\end{figure}

\begin{figure}[t]
\centering
\includegraphics[width=0.475\textwidth]{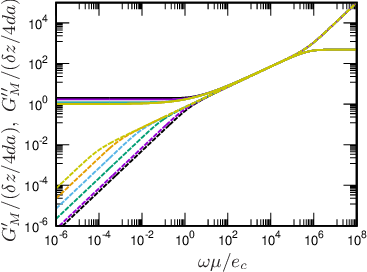}
\caption{\label{figref3con}
{Complex modulus at finite prestresses.}
$G_M^\ast/(\delta z/4da)$ is plotted as a function of $\omega \mu/e_c$.
Solid and dashed lines represent the storage modulus $G_M'$ and the loss modulus $G_M''$, respectively.
The excess contact number is $\delta z= 10^{-2}$.
The prestress is $(e_c -e)/e_c = 5\times 10^{-5}$~(yellow), $5\times 10^{-4}$~(orange), $5\times 10^{-3}$~(cyan), $5\times 10^{-2}$~(green), $5\times 10^{-1}$~(purple), and $1$~($e=0$, black).
}
\end{figure}

We also numerically solve Eqs.~(\ref{emteqcon1over}) and~(\ref{emteqcon2over}) and calculate $G_{M}^{\ast}$ using Eqs.~(\ref{eqkeff}) and~(\ref{eqgmast}).
Figures~\ref{figref2con} and~\ref{figref3con} present numerical solutions of $G_{M}^{\ast}$ as a function of $\omega \mu$.
In Fig.~\ref{figref2con}, we consider the case of zero prestress $e=0$ and show the dependence on $\delta z$, whereas, in Fig.~\ref{figref3con}, we set the contact number to be $\delta z = 10^{-2}$ and show the dependence on $e$.

Looking at Fig.~\ref{figref2con} for zero prestress, we confirm the scaling behaviors in Eq.~(\ref{eqconscale2}).
$G_{M}^{\ast}$ shows a crossover at $\omega \mu \approx e_c \propto \delta z^2$.
At $\omega \mu \ll e_c$, $G'_{M} \propto \omega^0$ is significantly larger than $G''_{M}$, indicating behavior characteristic of an elastic medium.
In contrast, at $\omega \mu \gg e_c$, both moduli are almost equal as $G'_{M} \approx G''_{M} \propto \sqrt{\omega \mu}$, suggesting behaviors of both an elastic and a viscous media.
These theoretical results explain numerical simulations~\cite{Tighe_2011,Baumgarten_2017,Hara_2023,Hara2_2023} and experimental observations~\cite{Mason_1995,Hebraud_2000,Gopal_2003,Kropka_2010,Nishizawa_2017,Hara2_2023}.
Note that in the highest frequency regime of $\omega \mu \gg 1$, $G'_{M} \approx 1$ and $G''_{M} \approx \omega \mu$ reflect the stiffness of single spring and the viscosity of viscous force, respectively~\cite{Hara2_2023}.

In addition, in Fig.~\ref{figref3con} at finite prestresses, we confirm the scaling behaviors in Eq.~(\ref{eqconscale}). 
Apart from the crossover at $\omega \mu \approx e_c$, an additional crossover occurs at $\omega \mu \approx \delta e$.
As a result, at $\delta e \ll \omega \mu \ll e_c$, we observe the scaling behavior characterized by $G'_{M} \propto \omega^0$ and $G''_{M,m} \propto \sqrt{\omega \mu}$.
This is the anomalous viscous loss~\cite{Liu_1996}, as has been observed in many experiments~\cite{Mason_1995,Cohen-Addad_1998,Hebraud_2000,Gopal_2003,Besson_2008,Kropka_2010,Krishan_2010,Gupta_2012,Basu_2014,Nishizawa_2017,Conley_2019,Hara2_2023}.
The anomalous viscous loss occurs at finite prestresses, which is thus caused by the prestress or the repulsive forces between particles. 
As $e$ approaches the stability limit $e_c$, the frequency range of anomalous viscous loss expands towards zero frequency, and at $e=e_c$,
\begin{equation} \label{eqanomalous}
\frac{G^\ast_M}{\delta z}
\propto
\left\{ 
\begin{aligned}
& 1 + i \sqrt{\frac{\omega \mu}{e_c}} & \left( \frac{\omega \mu}{e_c} \ll 1 \right), \\
& \sqrt{\frac{\omega \mu}{e_c}} + i \sqrt{\frac{\omega \mu}{e_c}} & \left( 1 \ll \frac{\omega \mu}{e_c} \ll \frac{1}{e_c} \right).
\end{aligned} 
\right.
\end{equation}
Therefore, the anomalous viscous loss is attributed to marginal stability induced by the prestress.

\section{Conclusion}
We have developed the EMT based on the random spring network model by integrating contact damping into the theory.
The theory explains the viscoelastic properties of soft jammed solids, such as foams, emulsions, and soft colloids, observed in experiments~\cite{Mason_1995,Cohen-Addad_1998,Hebraud_2000,Gopal_2003,Besson_2008,Kropka_2010,Krishan_2010,Gupta_2012,Basu_2014,Nishizawa_2017,Conley_2019,Hara2_2023}, demonstrating that strong damping plays a key role in determining their viscoelastic behaviors.
At lower frequencies, the system behaves like an elastic body with $G' \propto \omega^0 \gg G''$, whereas at higher frequencies, the system exhibits characteristics of both an elastic and a viscous media, with $G' \approx G'' \propto \sqrt{\omega}$.
Furthermore, the theory accounts for the anomalous viscous loss, characterized by $G' \propto \omega^0$ and $G'' \propto \sqrt{\omega}$, which extends towards zero frequency as the system approaches the marginally stable state.

An important result of the EMT is that the anomalous viscous loss is attributed to marginal stability.
Marginal stability plays a crucial role in amorphous systems~\cite{Muller_2015}.
In particular, the boson peak and the quasi-localized vibrations~\cite{Silbert_2005,Wyart_2005,Wyart2_2005,Charbonneau2016,Mizuno_2017,Shimada_2020} are consequences of the marginal stability, which is explained by the EMT~\cite{Wyart_2010,Degiuli_2014,Shimada2_2020,Shimada_2021}.
In the EMT, the boson peak corresponds to the non-Debye scaling law of the vDOS $D(\omega) \propto \omega^2$, which also persists to the zero frequency at the marginally stable state, as does the anomalous viscous loss.
Therefore, the boson peak and the anomalous viscous loss are linked through marginal stability, providing an explanation for the findings reported in Ref.~\cite{Hara2_2023}.

In this work, we examined contact damping and showed that both macrorheology and microrheology yield the same complex moduli.
However, for Stokes damping, there can be differences between the two measurements~\cite{Hara_2023}, which needs to be addressed in future research. 
Furthermore, we studied the overdamped dynamics, which are characteristic of soft jammed solids like foams, emulsions, and soft colloids.
In contrast, Refs.~\cite{Wyart_2010,Degiuli_2014} examined the opposite extreme case of zero damping, characteristic of hard jammed solids like structural glasses at low temperatures.
As discussed in Ref.~\cite{Mizuno_2024}, the level of damping plays a critical role in distinguishing between soft and hard jammed solids.
Intermediate levels of damping may help to explain material properties that fall between these extremes, such as the behavior of glasses at finite temperatures~\cite{Mizuno_2019,Mizuno_2020,Wang_2020}.
Finally, it would be interesting to integrate strong damping effects into other theoretical frameworks, including the HET with fluctuating elastic constants~\cite{schirmacher_2006,Schirmacher_2007,Marruzzo_2013,Kohler_2013,Caroli_2019,Kapteijns_2021,Mahajan_2021}, the random matrix approach~\cite{Conyuh_2021,Vogel_2023,Philipp_2024}, and other microscopic approaches~\cite{Damart_2017,Baggioli_2022,Szamel_2022}.

\section*{Acknowledgments}
This work was supported by JSPS KAKENHI Grant Numbers 22K03543, 23H04495, and 24H00192.

\appendix
\section{Asymptotic analysis}~\label{sec:asympto}
To find asymptotic solution in Eqs.~(\ref{emteqcon1over}) and~(\ref{emteqcon2over}), we use the asymptotic analysis reported in Ref.~\cite{Degiuli_2014} as follows.
We suppose $\delta z \ll 1$, and find the solution of ${k}^{\parallel}$ and $k^{\perp}$ in the form of
\begin{equation}
{k}^{\parallel} = {k}^{\parallel}_0 \delta z + o\left(\delta z \right),\qquad k^{\perp} = k^{\perp}_0 = {O}\left(\delta z^0 \right).
\end{equation}
As shown in Ref.~\cite{Degiuli_2014}, there is a critical value of the prestress $e$, denoted by $e_c$, as
\begin{equation}
e_c = e_1 \delta z^2 + o\left(\delta z^2 \right),
\end{equation}
and $e$ is expressed as $e = e^\prime e_c$ with $0 \le e^\prime \le 1$, thus $e = {O}\left(\delta z^2 \right)$.
We also suppose $\omega \mu = {O}\left(\delta z^2 \right)$.
Then, $\widetilde{k}$ and $\widetilde{e}$ in Eq.~(\ref{eqtildeke}) are estimated as
\begin{equation}
\begin{aligned}
\widetilde{k} &= 1 - i {\omega \mu} = 1 + {O}\left(\delta z^2 \right), \\
\widetilde{e} &= \frac{e + i \omega \mu}{ 1 - i \omega \mu} = \widetilde{e}^\prime e_c + {O}\left(\delta z^4 \right),
\end{aligned}
\end{equation}
where $\widetilde{e}^\prime = e^\prime + i \omega \mu/e_c = {O}\left(\delta z^0 \right)$.
The effective stiffness $k_\text{eff}$ in Eq.~(\ref{eqkeff}) is evaluated as
\begin{equation}
k_\text{eff} = {k}^{\parallel}_0 \delta z + o\left(\delta z \right).
\end{equation}
Under these assumptions, we find the solution in Eqs.~(\ref{emteqcon1over}) and~(\ref{emteqcon2over}).

From Eq.~(\ref{emteqcon2over}) for ${G}^\parallel$, we obtain
\begin{equation} \label{eqgparacon1}
\frac{z_0}{2d} k_\text{eff} {G}^\parallel = 1,
\end{equation}
whereas, from Eq.~(\ref{emteqcon1over}) for ${G}^\parallel$, we obtain
\begin{equation} \label{eqgparacon2}
\frac{z_0}{2d} k_\text{eff} {G}^\parallel = 1 +  \left( -a {k}_0^\parallel - \frac{2 \widetilde{e}^\prime e_1 {k}_0^\perp}{{k}_0^\parallel} + \frac{1}{2d} \right) \delta z + {o}\left(\delta z \right).
\end{equation}
By equalizing Eqs.~(\ref{eqgparacon1}) and~(\ref{eqgparacon2}), we obtain
\begin{equation} \label{eqgparacon3}
{k}_0^\parallel = \frac{1}{4da} \left( 1 \pm \sqrt{ 1 - 32 d^2 a \widetilde{e}^\prime e_1 {k}_0^\perp} \right).
\end{equation}
In addition, from Eq.~(\ref{emteqcon1over}) for ${G}^\perp$, we obtain
\begin{equation}
\frac{z_0}{2d} k_\text{eff} {G}^\perp = \frac{z_0}{2d} \frac{k_0^\parallel}{ \widetilde{e}^\prime e_1 k_0^\perp ({k}_0^\perp-1)} \frac{1}{\delta z}\left( \frac{z}{z_0} - k_0^\perp \right),
\end{equation}
which should be ${O}\left(\delta z^0 \right)$, thus 
\begin{equation} \label{eqk0perp}
{k}_0^\perp = \frac{2d}{z_0}.
\end{equation}

Gathering the above results, we obtain
\begin{equation} \label{appeqkeff}
{k}_\text{eff} = \left( 1 \pm \sqrt{ 1 - 32 d^2 a \widetilde{e}^\prime e_1 \frac{2d}{z_0}} \right)\frac{\delta z}{4da} + {o}\left(\delta z \right).
\end{equation}
By setting $\omega \mu =0$, we obtain the static value ${k}_{\text{eff}0}$ as
\begin{equation} \label{staticeff}
{k}_{\text{eff}0} = \left( 1 \pm \sqrt{ 1 - 32 d^2 a {e}^\prime e_1 \frac{2d}{z_0}} \right)\frac{\delta z}{4da} + {o}\left(\delta z \right).
\end{equation}
Since ${k}_{\text{eff}0}$ should be real at $e^\prime < 1$ and becomes complex at $e^\prime > 1$, we obtain
\begin{equation}
e_1 = \frac{1}{32 d^2 a} \frac{z_0}{2d},
\end{equation}
and $e_c$ in Eq.~(\ref{eqec}).
Thus, Eq.~(\ref{appeqkeff}) becomes
\begin{equation} \label{appeqkeff2}
{k}_\text{eff} = \left( 1 \pm \sqrt{ 1 - e^\prime - i \frac{\omega \mu}{e_c} } \right)\frac{\delta z}{4da} + {o}\left(\delta z \right).
\end{equation}

We finally determine the sign in Eq.~(\ref{appeqkeff2}).
At low frequencies $\omega \mu /e_c \ll 1 - e^\prime$, Eq.~(\ref{appeqkeff2}) becomes
\begin{equation}
{k}_\text{eff} \approx \left[ 1 \pm \sqrt{1- e^\prime}\left(1 - i \frac{\omega \mu}{2e_c (1-e^\prime)} \right) \right] \frac{\delta z}{4da},
\end{equation}
where the sign should be positive, as the real and imaginary parts of ${k}_\text{eff}$ should be positive and negative, respectively.
On the other hand, at high frequencies $\omega \mu /e_c \gg 1- e^\prime$, Eq.~(\ref{appeqkeff2}) becomes
\begin{equation}
{k}_\text{eff} \approx \left[ 1 \pm (1 - i) \sqrt{ \frac{\omega \mu}{2e_c} } \right] \frac{\delta z}{4da},
\end{equation}
and the sign should still be positive.
Therefore, we write Eq.~(\ref{appeqkeff2}) as
\begin{equation}
{k}_\text{eff} = \left( 1 + \sqrt{ 1 - e^\prime - i \frac{\omega \mu}{e_c} } \right)\frac{\delta z}{4da} + {o}\left(\delta z \right),
\end{equation}
which results in $G_{M}^{\ast}$ in Eq.~(\ref{mainemaccon}).

\bibliographystyle{apsrev4-2}
\bibliography{reference}

\end{document}